\pdfoutput=1

\documentclass[12pt,a4paper]{article}

\usepackage{ifthen} 
\newboolean{pdflatex}
\setboolean{pdflatex}{true} 

\newboolean{articletitles}
\setboolean{articletitles}{true} 

\newboolean{uprightparticles}
\setboolean{uprightparticles}{false} 

\newboolean{inbibliography}
\setboolean{inbibliography}{false} 
\usepackage{multirow}

\def\paperauthors{LHCb collaboration} 
\def\paperasciititle{Open charm production and asymmetry in pNe collisions at SQRT(sNN)=68.5 GeV} 
\def\papertitle{Open charm production and asymmetry in $p$Ne collisions at $\sqrt{s_{\scriptscriptstyle\rm NN}} =$ 68.5\gev} 
\def\paperkeywords{{High Energy Physics}, {LHCb}} 
\def\papercopyright{2021 CERN for the benefit of the LHCb collaboration}
\def\paperlicence{CC-BY-4.0}
\def\paperlicenceurl{https://creativecommons.org/licenses/by/4.0/}

\usepackage[top=1in, bottom=1.25in, left=1in, right=1in]{geometry}

\usepackage{microtype}
\usepackage{lineno}  
\usepackage{xspace} 
\usepackage{caption} 

\usepackage{graphicx}  
\usepackage{color}
\usepackage{colortbl}
\graphicspath{{./figs/}} 

\usepackage{amsmath} 
\usepackage{amssymb}
\usepackage{amsfonts}
\usepackage{upgreek} 

\newcommand*\patchAmsMathEnvironmentForLineno[1]{%
\expandafter\let\csname old#1\expandafter\endcsname\csname #1\endcsname
\expandafter\let\csname oldend#1\expandafter\endcsname\csname
end#1\endcsname
 \renewenvironment{#1}%
   {\linenomath\csname old#1\endcsname}%
   {\csname oldend#1\endcsname\endlinenomath}%
}
\newcommand*\patchBothAmsMathEnvironmentsForLineno[1]{%
  \patchAmsMathEnvironmentForLineno{#1}%
  \patchAmsMathEnvironmentForLineno{#1*}%
}
\AtBeginDocument{%
\patchBothAmsMathEnvironmentsForLineno{equation}%
\patchBothAmsMathEnvironmentsForLineno{align}%
\patchBothAmsMathEnvironmentsForLineno{flalign}%
\patchBothAmsMathEnvironmentsForLineno{alignat}%
\patchBothAmsMathEnvironmentsForLineno{gather}%
\patchBothAmsMathEnvironmentsForLineno{multline}%
\patchBothAmsMathEnvironmentsForLineno{eqnarray}%
}

\usepackage{hyperxmp}

\usepackage[pdftex,
            pdfauthor={\paperauthors},
            pdftitle={\paperasciititle},
            pdfkeywords={\paperkeywords},
            pdfcopyright={Copyright (C) \papercopyright},
            pdflicenseurl={\paperlicenceurl}]{hyperref}

\usepackage[all]{hypcap} 

\usepackage{xspace} 
\usepackage{upgreek}

\def\lhcb   {\mbox{LHCb}\xspace}

\def\MagUp {\mbox{\em Mag\kern -0.05em Up}\xspace}

\ifthenelse{\boolean{uprightparticles}}%
{

 \def\Ppi         {\ensuremath{\uppi}\xspace}

 \def\PDelta      {\ensuremath{\Delta}\xspace}                 
 \def\PXi         {\ensuremath{\Xi}\xspace}                 
 \def\PLambda     {\ensuremath{\Lambda}\xspace}                 
 \def\PSigma      {\ensuremath{\Sigma}\xspace}                 
 \def\POmega      {\ensuremath{\Omega}\xspace}                 
 \def\PUpsilon    {\ensuremath{\Upsilon}\xspace}
 \let\oldPi\Pi
 \def\PPi         {\ensuremath{\oldPi}\xspace}

 \def\PB      {\ensuremath{\mathrm{B}}\xspace}                 
                  
 \def\PD      {\ensuremath{\mathrm{D}}\xspace}

 \def\PK      {\ensuremath{\mathrm{K}}\xspace}

 \def\Pb      {\ensuremath{\mathrm{b}}\xspace}                 
 \def\Pc      {\ensuremath{\mathrm{c}}\xspace}

 \def\Pi      {\ensuremath{\mathrm{i}}\xspace}

 \def\Ps      {\ensuremath{\mathrm{s}}\xspace}                 
                  
 \def\Pu      {\ensuremath{\mathrm{u}}\xspace}

 \def\thebaroffset{0.0em}
}
{

 \def\Ppi         {\ensuremath{\pi}\xspace}

 \mathchardef\PDelta="7101
 \mathchardef\PXi="7104
 \mathchardef\PLambda="7103
 \mathchardef\PSigma="7106
 \mathchardef\POmega="710A
 \mathchardef\PUpsilon="7107
 \mathchardef\PPi="7105
                  
 \def\PB      {\ensuremath{B}\xspace}                 
                  
 \def\PD      {\ensuremath{D}\xspace}

 \def\PK      {\ensuremath{K}\xspace}

 \def\Pb      {\ensuremath{b}\xspace}                 
 \def\Pc      {\ensuremath{c}\xspace}

 \def\Pi      {\ensuremath{i}\xspace}

 \def\Ps      {\ensuremath{s}\xspace}                 
                  
 \def\Pu      {\ensuremath{u}\xspace}

 \def\thebaroffset{0.18em}
}
\newcommand{\offsetoverline}[2][\thebaroffset]{\kern #1\overline{\kern -#1 #2}}%

\makeatletter
\ifcase \@ptsize \relax
  \newcommand{\miniscule}{\@setfontsize\miniscule{4}{5}}
\or
  \newcommand{\miniscule}{\@setfontsize\miniscule{5}{6}}
\or
  \newcommand{\miniscule}{\@setfontsize\miniscule{5}{6}}
\fi
\makeatother

\DeclareRobustCommand{\optbar}[1]{\shortstack{{\miniscule (\rule[.5ex]{1.25em}{.18mm})}
  \\ [-.7ex] $#1$}}

\def\uquark    {{\ensuremath{\Pu}}\xspace}
\def\uquarkbar {{\ensuremath{\overline \uquark}}\xspace}

\def\squark    {{\ensuremath{\Ps}}\xspace}

\def\cquark    {{\ensuremath{\Pc}}\xspace}
\def\cquarkbar {{\ensuremath{\overline \cquark}}\xspace}
\def\ccbar     {{\ensuremath{\cquark\cquarkbar}}\xspace}
\def\bquark    {{\ensuremath{\Pb}}\xspace}

\def\pion   {{\ensuremath{\Ppi}}\xspace}

\def\pip    {{\ensuremath{\pion^+}}\xspace}
\def\pim    {{\ensuremath{\pion^-}}\xspace}

\def\kaon    {{\ensuremath{\PK}}\xspace}

\def\KorKbar {\kern \thebaroffset\optbar{\kern -\thebaroffset \PK}{}\xspace}

\def\Kp      {{\ensuremath{\kaon^+}}\xspace}
\def\Km      {{\ensuremath{\kaon^-}}\xspace}

\def\Dbar    {{\ensuremath{\offsetoverline{\PD}}}\xspace}
\def\D       {{\ensuremath{\PD}}\xspace}

\def\DorDbar {\kern \thebaroffset\optbar{\kern -\thebaroffset \PD}\xspace}
\def\Dz      {{\ensuremath{\D^0}}\xspace}
\def\Dzb     {{\ensuremath{\Dbar{}^0}}\xspace}
\def\Dp      {{\ensuremath{\D^+}}\xspace}
\def\Dm      {{\ensuremath{\D^-}}\xspace}

\def\DpDm    {\ensuremath{\Dp {\kern -0.16em \Dm}}\xspace}
\def\Dstar   {{\ensuremath{\D^*}}\xspace}

\def\B       {{\ensuremath{\PB}}\xspace}

\def\BorBbar {\kern \thebaroffset\optbar{\kern -\thebaroffset \PB}\xspace}

\def\Bd      {{\ensuremath{\B^0}}\xspace}

\def\BdorBdbar {\kern \thebaroffset\optbar{\kern -\thebaroffset \Bd}\xspace}

\def\Bs      {{\ensuremath{\B^0_\squark}}\xspace}

\def\BsorBsbar {\kern \thebaroffset\optbar{\kern -\thebaroffset \Bs}\xspace}

\def\Y#1S{\ensuremath{\PUpsilon{(#1S)}}\xspace}

\def\LorLbar     {\kern \thebaroffset\optbar{\kern -\thebaroffset \PLambda}\xspace}

\def\BF         {{\ensuremath{\mathcal{B}}}\xspace}

\def\to                 {\ensuremath{\rightarrow}\xspace}

\def\AT#1     {\ensuremath{A_{\mathrm{T}}^{#1}}\xspace}           

\def\C#1      {\ensuremath{\mathcal{C}_{#1}}\xspace}                       
\def\Cp#1     {\ensuremath{\mathcal{C}_{#1}^{'}}\xspace}                    
\def\Ceff#1   {\ensuremath{\mathcal{C}_{#1}^{\mathrm{(eff)}}}\xspace}        
\def\Cpeff#1  {\ensuremath{\mathcal{C}_{#1}^{'\mathrm{(eff)}}}\xspace}       
\def\Ope#1    {\ensuremath{\mathcal{O}_{#1}}\xspace}                       
\def\Opep#1   {\ensuremath{\mathcal{O}_{#1}^{'}}\xspace}                    

\newcommand{\aunit}[1]{\ensuremath{\text{\,#1}}}       

\newcommand{\tev}{\aunit{Te\kern -0.1em V}\xspace}
\newcommand{\gev}{\aunit{Ge\kern -0.1em V}\xspace}
\newcommand{\mev}{\aunit{Me\kern -0.1em V}\xspace}
\newcommand{\kev}{\aunit{ke\kern -0.1em V}\xspace}
\newcommand{\ev}{\aunit{e\kern -0.1em V}\xspace}
 
\newcommand{\mevc}{\ensuremath{\aunit{Me\kern -0.1em V\!/}c}\xspace}
\newcommand{\gevc}{\ensuremath{\aunit{Ge\kern -0.1em V\!/}c}\xspace}
\newcommand{\mevcc}{\ensuremath{\aunit{Me\kern -0.1em V\!/}c^2}\xspace}
\newcommand{\gevcc}{\ensuremath{\aunit{Ge\kern -0.1em V\!/}c^2}\xspace}

\newcommand{\stat}{\aunit{(stat.)}\xspace}
\newcommand{\syst}{\aunit{(syst.)}\xspace}

\def\gsim{{~\raise.15em\hbox{$>$}\kern-.85em
          \lower.35em\hbox{$\sim$}~}\xspace}
\def\lsim{{~\raise.15em\hbox{$<$}\kern-.85em
          \lower.35em\hbox{$\sim$}~}\xspace}

\def\sqs   {\ensuremath{\protect\sqrt{s}}\xspace}
\def\sqsnn {\ensuremath{\protect\sqrt{s_{\scriptscriptstyle\text{NN}}}}\xspace}
\def\pt         {\ensuremath{p_{\mathrm{T}}}\xspace}

\def\evtgen     {\mbox{\textsc{EvtGen}}\xspace}

\def\geant      {\mbox{\textsc{Geant4}}\xspace}

\def\photos     {\mbox{\textsc{Photos}}\xspace}

\def\pythia     {\mbox{\textsc{Pythia}}\xspace}

\def\tell1  {TELL1\xspace}
\def\ukl1   {UKL1\xspace}

\newcommand{\lhcborcid}[1]{\href{https://orcid.org/#1}{\hspace*{0.1em}\raisebox{-0.45ex}{\includegraphics[width=1em]{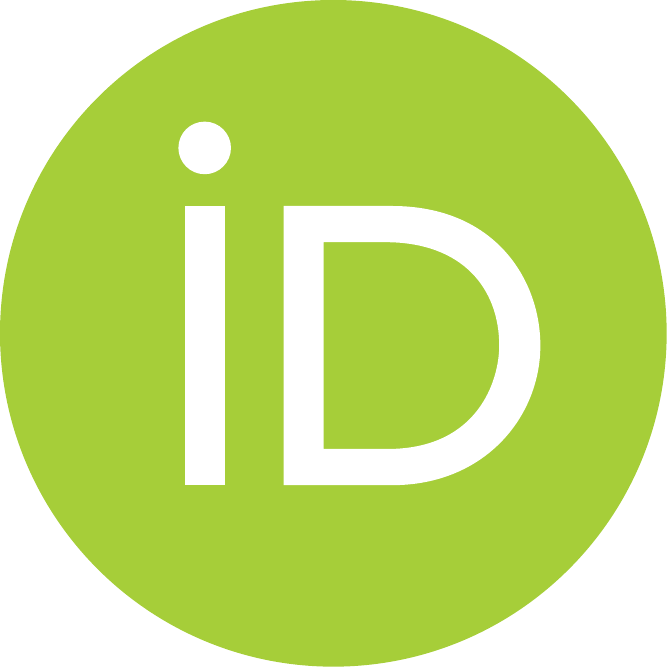}}}}

\def\epos      {\mbox{\textsc{Epos}}\xspace}
\newcommand{\pp}{\ensuremath{pp}\xspace}
\newcommand{\pHe}{\ensuremath{p\text{He}}\xspace}

\newcommand{\pNe}{\ensuremath{p\text{Ne}}\xspace}

\usepackage{cite} 
\usepackage{mciteplus}

\usepackage{longtable} 

\begin{document}

\renewcommand{\thefootnote}{\fnsymbol{footnote}}
\setcounter{footnote}{1}

\begin{titlepage}
\pagenumbering{roman}

\vspace*{-1.5cm}
\centerline{\large EUROPEAN ORGANIZATION FOR NUCLEAR RESEARCH (CERN)}
\vspace*{1.5cm}
\noindent
\begin{tabular*}{\linewidth}{lc@{\extracolsep{\fill}}r@{\extracolsep{0pt}}}
\ifthenelse{\boolean{pdflatex}}
{\vspace*{-1.5cm}\mbox{\!\!\!\includegraphics[width=.14\textwidth]{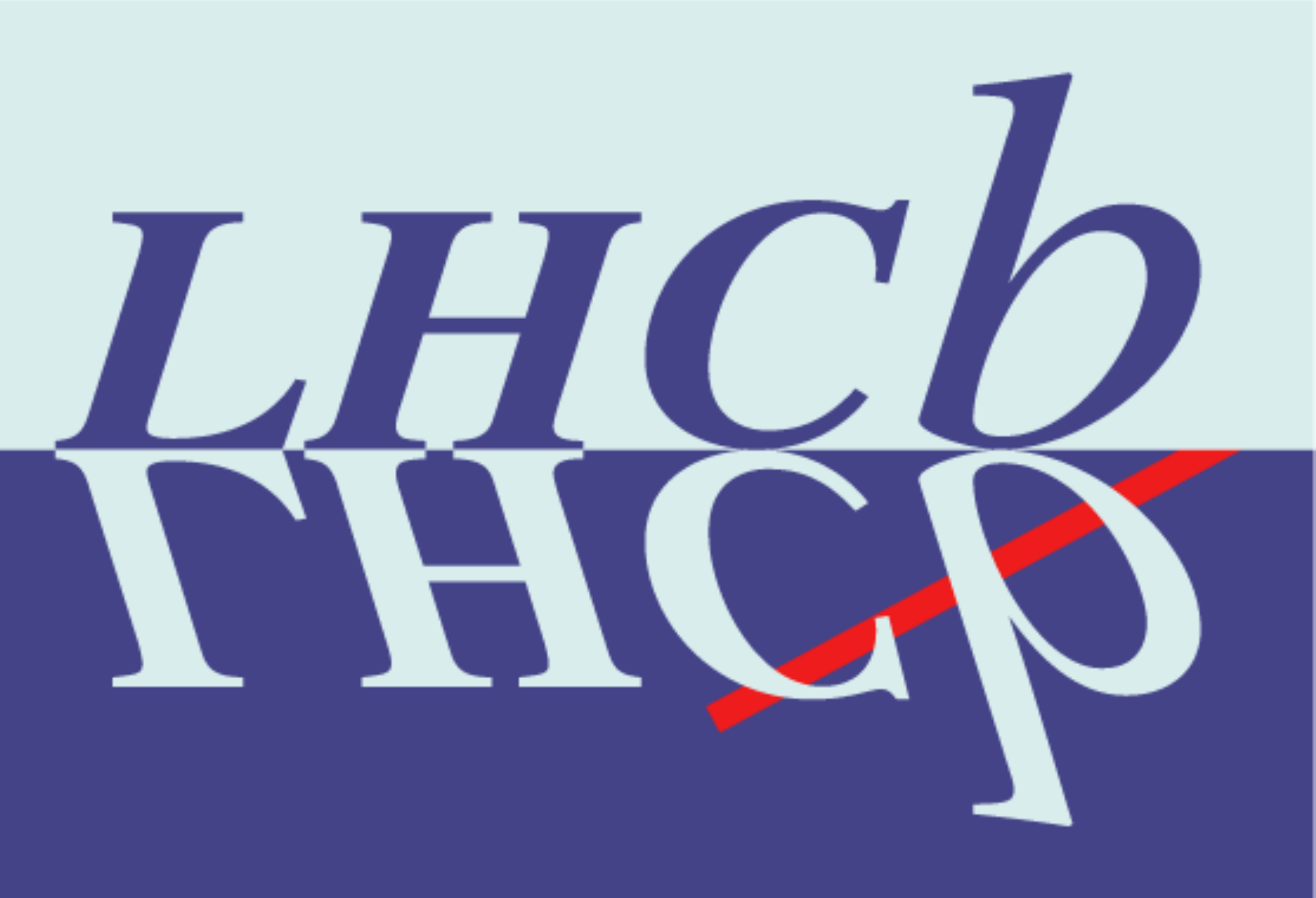}} & &}%
{\vspace*{-1.2cm}\mbox{\!\!\!\includegraphics[width=.12\textwidth]{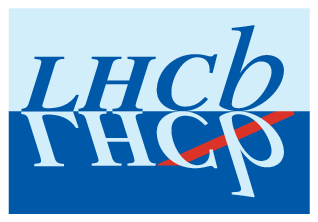}} & &}%
\\
 & & CERN-EP-2022-217 \\  
 & & LHCb-PAPER-2022-015 \\  
 & & August 10, 2023 \\ 
 & & \\
\end{tabular*}

\vspace*{4.0cm}

{\normalfont\bfseries\boldmath\huge
\begin{center}
  \papertitle 
\end{center}
}

\vspace*{1.0cm}

\begin{center}
\paperauthors\footnote{Authors are listed at the end of this paper.}
\end{center}

\vspace{\fill}

\begin{abstract}
  \noindent
A measurement of \Dz meson production by the \lhcb experiment in its fixed-target configuration is presented. The production of \Dz mesons is studied with a beam of 2.5\tev protons colliding on a gaseous neon target at rest, corresponding to a nucleon-nucleon centre-of-mass energy of $\sqsnn = 68.5\gev$. The sum of the \Dz and \Dzb production cross-section in \pNe collisions in the centre-of-mass rapidity range $y^{\star}\in [-2.29, 0]$ is found to be $\sigma_{D^{0}}^{y^\star \in [-2.29, 0]} = 48.2 \pm 0.3 \pm 4.5 \,\upmu\textrm{b/nucleon}$ where the first uncertainty is statistical and the second is systematic. The $\Dz-\Dzb$ production asymmetry is also evaluated and suggests a trend towards negative values at large negative $y^{\star}$. The considered models do not account precisely for all the features observed in the LHCb data, but theoretical predictions including 1$\%$ intrinsic charm and 10$\%$ recombination contributions better describe the data than the other models considered.  
\end{abstract}

\vspace*{2.0cm}

\begin{center}
Published in Eur.~Phys.~J.~\textbf{C83} (2023) 541.
\end{center}

\vspace{\fill}

{\footnotesize 
\centerline{\copyright~\papercopyright. \href{\paperlicenceurl}{\paperlicence} licence.}}
\vspace*{2mm}

\end{titlepage}


\newpage
\setcounter{page}{2}
\mbox{~}

\cleardoublepage

\renewcommand{\thefootnote}{\arabic{footnote}}
\setcounter{footnote}{0}


\pagestyle{plain} 
\setcounter{page}{1}
\pagenumbering{arabic}


The production of \Dz charm mesons is sensitive to non-perturbative aspects of Quantum Chromodynamics (QCD).
As it reflects a large fraction of the overall charm quark production, \Dz meson production can serve as a reference for the study of the modification of hidden charm (\ccbar bound states) production in proton-nucleus and nucleus-nucleus collisions, due to the so-called cold nuclear matter (CNM) and hot and dense matter (HDM) effects  \cite{Satz_1994}.    
Additionally, the study of \Dz and \Dzb mesons, \cquark\uquarkbar and \cquarkbar\uquark bound states, may bring new insight on the intrinsic charm content of the nucleon \cite{maciula2021impact}, as already performed with the proton-helium (\pHe) data collected with \lhcb at $\sqsnn=86.6\gev$ \cite{D0pHeLHCb}, complementing the results obtained in proton-proton (\pp) collisions at $\sqsnn=13\tev$ via the measurement of $Z+\cquark$ jets \cite{Z0jetLHCb,NNPDF}. Moreover, in collisions involving a high Bjorken-$x$ parton the charm quarks may recombine with valence quarks from the initial hadrons and lead to an asymmetry between the \Dz and \Dzb production cross-sections \cite{MS, E791, Braaten}. 

In this paper, a measurement of \Dz production in fixed-target proton-neon (\pNe) collisions at the LHC is presented. The sum of the \Dz and \Dzb production, and the $\Dz - \Dzb$ production asymmetry, are studied in collisions of protons with energy of 2.5\tev incident on neon nuclei at rest, corresponding to a nucleon-nucleon centre-of-mass energy of $\sqsnn=68.5\gev$. These measurements are performed in the negative rapidity hemisphere in the centre-of-mass frame where the proton beam and neon target have positive and negative rapidity $y^\star$, respectively.

The \lhcb detector~\cite{Alves:2008zz,LHCb-DP-2014-002} is a single-arm forward spectrometer designed for the study of particles containing \cquark or \bquark quarks, covering the pseudorapidity range $2 < \eta < 5$ in the laboratory frame.
The detector elements that are particularly relevant to this analysis are: the silicon-strip vertex locator (VELO) surrounding the interaction region that allows \cquark and \bquark hadrons to be identified from their characteristic flight distance; a tracking system that provides a measurement of the momentum of charged particles; two ring-imaging Cherenkov detectors that are able to discriminate between different species of charged hadrons; a calorimeter system consisting of scintillating-pad and preshower detectors, electromagnetic and hadronic calorimeters; and a muon detector composed of alternating layers of iron and multiwire proportional chambers. The system for measuring the overlap with gas (SMOG)~\cite{LHCb-PAPER-2014-047} enables the injection of gases with a pressure of $O(10^{-7})$ mbar in the beam pipe section inside the VELO, allowing \lhcb to operate as a fixed-target experiment. SMOG allows the injection of noble gases and therefore gives the unique opportunity to study nucleus-nucleus and proton-nucleus collisions for various gaseous targets. Due to the boost induced by the high-energy proton beam, which travels along the positive-$z$ direction, the \lhcb acceptance covers the negative rapidity hemisphere in the centre-of-mass system of the reaction  $y^\star \in$  $[-2.29, 0]$. 

The data samples correspond to a collider configuration in which proton bunches moving towards the detector do not cross any bunch moving in the opposite direction at the nominal \pp interaction point. Standard \pp collision events were also collected concurrently with the \pNe collision data.

Events are selected  by the two-stage trigger system \cite{trigger}. The first level is implemented in hardware, while the second is a software trigger. The hardware trigger requires a minimum deposited energy of 7.8 \gev in the calorimeter  for the $\Dz \to K^- \pi^+$ selection.\footnote{Inclusion of charge-conjugate processes is implied throughout, unless otherwise said.} The software trigger requires a well-reconstructed \Dz decay vertex formed by clearly identified kaon and pion tracks, both of which are required to have a transverse momentum greater than 250 \mevc and a combined invariant mass in the range [1784, 1944]~\mevcc.

Special care is taken to suppress residual \pp collisions induced by debunched protons. The PV must lie in the fiducial region $z_\textrm{PV}\in[-200,-100]\cup[100,150]$~mm, where high reconstruction efficiencies are achieved and calibration samples are available. Here $z_\textrm{PV}$ is the reconstructed position along the beam axis. The region $-100 < z_\textrm{PV} < 100$~mm, where most of the residual \pp collisions occur, is excluded.
A veto is imposed on events with activity in the backward region, with respect to the proton beam direction, based on the number of hits in the VELO stations upstream of the interaction point. 

The offline selection of \Dz candidates is similar to that used in Ref~\cite{LHCb-PAPER-2018-023}.
Events must contain a primary vertex with at least four tracks reconstructed in the VELO detector.  The kaon and pion originating from the \Dz decay are required to be of good quality and to come from a common displaced vertex relatively to the associated PV position to which the \Dz candidate has the smallest impact parameter. Tight requirements are set on the kaon and pion particle
identification information. The \Dz candidates are required to have a proper decay time greater than 0.5 ps. The measurements are performed in the range of \Dz transverse momentum
$\pt < 8$ \gevc and  $2.0 < y < 4.29$, where $y$ is the rapidity in the laboratory frame.

The detection efficiencies are determined using samples of simulated \pNe collisions. In the simulation, \Dz mesons are generated in simulated proton-proton collisions using \pythia 8 \cite{Pythia8}, which has implemented the Lund string fragmentation model \cite{Lund}, with a specific \lhcb configuration \cite{1742-6596-331-3-032047} and with colliding-proton beam momenta equal to the momenta per nucleon of the beam and target in the centre-of-mass frame. The  decays are generated by \evtgen \cite{EvtGen}, in which final-state radiation is handled by \photos \cite{PHOTOS}. The four-momenta of the \Dz decay products are embedded into \pNe minimum bias events that are produced with the \epos event generator \cite{Pierog:2013ria}  using beam parameters obtained from the data. The event obtained after embedding is then boosted to the laboratory frame to get the fixed-target configuration. Decays of hadronic particles generated with \epos are also implemented with \evtgen. The interaction of the generated particles with the detector and its response, are handled with the \geant toolkit \cite{geant2, geant4} as described in Ref. \cite{gauss}. After reconstruction, the simulated events are assigned weights to ensure that the VELO cluster multiplicity distribution matches the distribution in the data.
\begin{figure}[t]
\begin{center}
\includegraphics[width=0.49\textwidth]{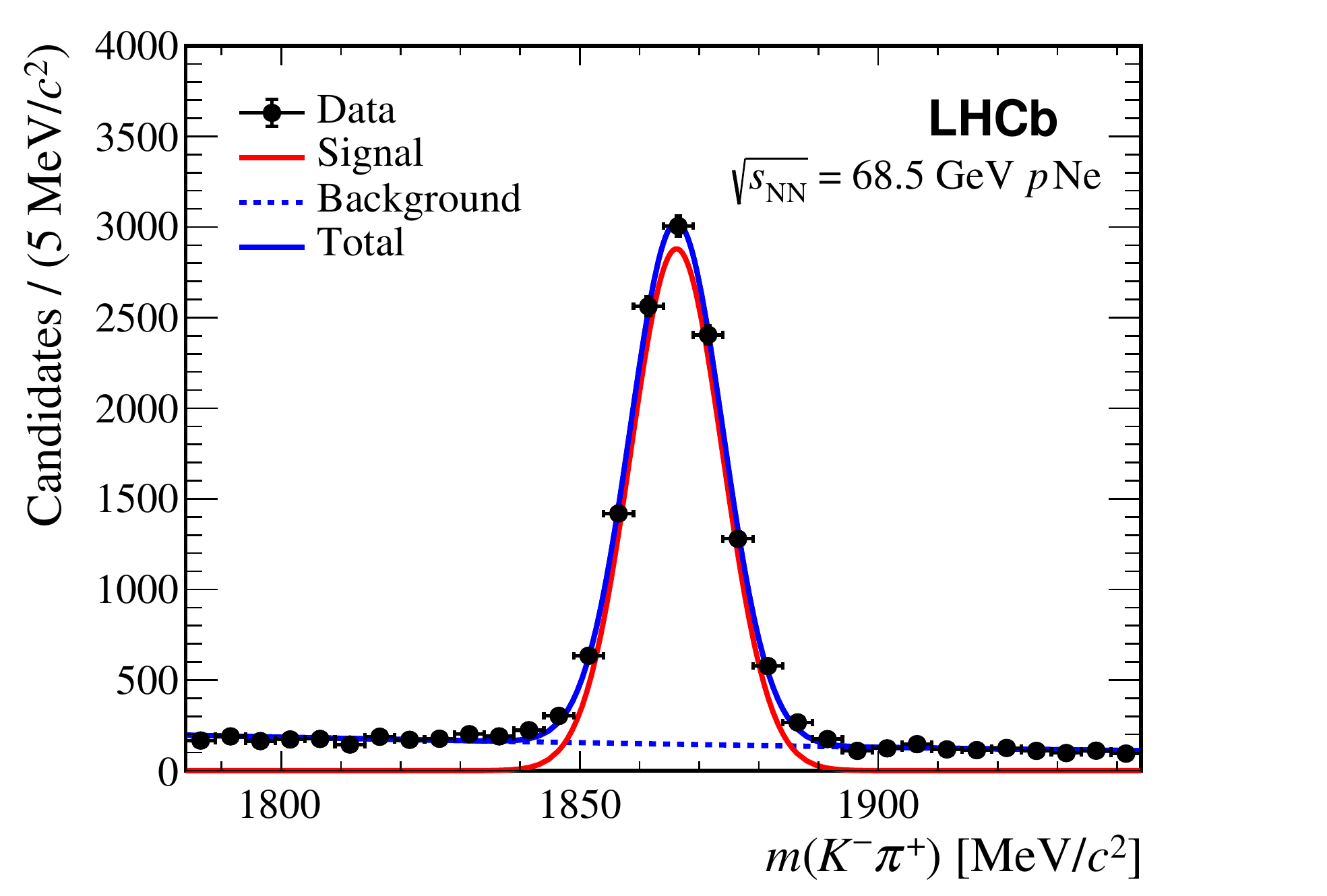}
\includegraphics[width=0.49\textwidth]{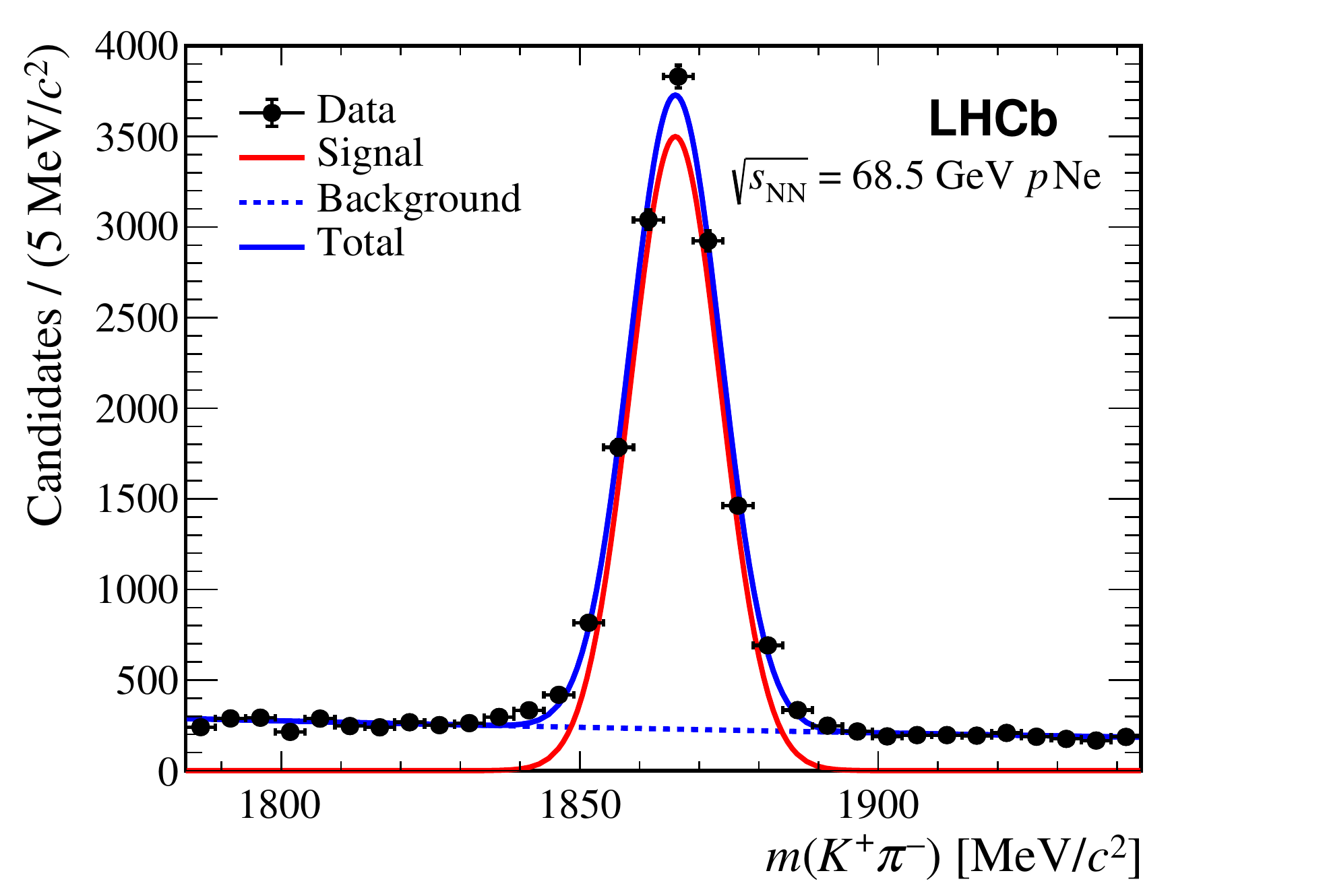}
\caption{Invariant mass distributions of (left) \Km\pip  and (right) \Kp\pim  candidates. The data are overlaid with the result of the fit. }
\label{fig::mass}
\end{center}
\end{figure}
The \Dz signal yields are obtained from extended unbinned maximum likelihood fits to their mass distributions. The signals are described by Gaussian functions while the background contributions are described by exponential functions. Figure \ref{fig::mass} shows the \Km\pip  and \Kp\pim invariant mass distributions obtained after all selection criteria are applied to the entire \pNe data set, with the fit functions superimposed. Additionally, the signal yields are determined in intervals of $\pt$ with edges (0, 0.5, 1, 1.5, 2, 3, 8 \gevc) and $y^{\star}$ with edges (-2.29, -1.5, -1.0, -0.5, 0). These yields are corrected for the total efficiencies, which account for the geometrical acceptance of the detector, and the efficiencies of the trigger, event selection including the \Dz selection, PV and track reconstruction, and particle identification. Particle identification \cite{LHCb-PUB-2016-021} and tracking \cite{tracking} efficiencies are obtained from control samples of \pp collision data. All other efficiencies are determined using samples of simulated data.

Several sources of systematic uncertainties are considered, affecting either the determination of the signal yields or the total efficiencies. They are summarised in Table~\ref{tab:systematics} separately for contributions that are correlated and uncorrelated between the kinematic intervals.
\begin{table}[t]
\caption{\small \label{tab:systematics} Systematic and statistical uncertainties on the signal yields. Systematic uncertainties correlated between $y^\star$ or \pt bins affect all measurements by the same relative amount. Ranges denote the minimum and the maximum values among the $y^\star$ or \pt bins.}
\begin{center}
{\small
\begin{tabular}{@{}lc@{}}
\hline
{\textbf{Systematic uncertainties}} & \\
\hline
{\textbf{Correlated between bins}} &   \\
Signal determination                 & 2.3$\%$ \\
Tracking efficiency                  & 3.0$\%$ \\
Particle identification efficiency   & 3.6$\%$ \\
PV reconstruction efficiency & 4.1$\%$ \\
Neon purity & 1.2$\%$ \\
Luminosity                          & 6.5$\%$ \\
{\textbf{Uncorrelated between bins}} &\\
Signal $\&$ background model        & [0.3, 5.6]$\%$  \\
Simulation sample                    & [0.7, 8.0]$\%$ \\ 
\hline
{\textbf{Total correlated uncertainty}} & 9.3$\%$ \\
{\textbf{Total uncorrelated uncertainty}} & [1.1, 9.2]$\%$ \\
{\textbf{Total statistical uncertainty}} & [2.3, 7.7]$\%$ \\ \hline
\end{tabular} 
}
\end{center}
\end{table}
Systematic uncertainties on the signal determination include the contribution from \bquark-hadron decays and the maximum contamination from residual \pp collisions. More precisely, the fraction of signal from \bquark-hadrons, included in the extracted yield and determined through the study of the impact parameter distribution of the \Dz candidates with respect to the primary vertex, is 1.1$\%$. Contributions from residual \pp collisions are estimated using samples of pure \pNe collisions and pure \pp collisions, both samples being collected using dedicated LHC beam configurations.  These contributions result in a total 2.3$\%$ uncertainty assigned to the signal determination. Since the tracking and particle identification efficiencies are determined using \pp control samples, the differences between the track multiplicity in \pNe and \pp collisions are considered as systematic uncertainties. Overall, systematic uncertainties of 3.0$\%$ and 3.6$\%$ are assigned due to tracking and particle identification efficiencies, respectively. The PV reconstruction systematic uncertainty is determined by considering the variation of the efficiency over the whole $z_\textrm{PV}$ range and  the difference between the simulation and a data-driven approach to evaluate the PV efficiency.
Possible contamination from collisions between the beam and atoms different from neon is quantified  using data samples recorded with no neon injection, resulting in an upper limit of 1.2$\%$, identified as the neon purity uncertainty. Two uncorrelated systematic uncertainties are also considered. First, the uncertainty related to the mass fit is evaluated using alternative models for signal and background shapes that reproduce the mass shapes equally well.
Another source of uncertainty is associated with the accuracy of the simulation used to compute the acceptances and efficiencies. 
This uncertainty accounts for the finite size of the simulation	sample and observed discrepancies between the data and the simulation, in the rapidity and \pt distributions.

The integrated luminosity is determined to be ${\cal{L}}_{p\textrm{Ne}} = 21.7 \pm 1.4 $ nb$^{-1}$ from the yield of electrons elastically scattering off the target atoms as presented in Ref. \cite{antiproton}.  
The measured \Dz and \Dzb  cross-section per target nucleon within $y^\star \in [-2.29, 0]$, after taking into account the known  branching fraction of $D^{0} \to \Km \pip$ \cite{pdg}, is
\begin{equation*}
      \sigma_{\Dz}^{y^\star \in [-2.29, 0]}  =  \frac{Y_{\Dz\to\Km\pip}}{\BF_{\Dz\to\Km\pip}\times\varepsilon_\Dz\times{\cal{L}}_{p\textrm{Ne}}} = 48.2 \pm 0.3 \stat \pm  4.5  \syst \, \upmu \textrm{b/nucleon},
\end{equation*}
where $Y_{\Dz\to\Km\pip}$, $\BF_{\Dz\to\Km\pip}$ and $\varepsilon_\Dz$ are \Dz yield, branching fraction and total efficiency respectively. No correction is applied to account for the small fraction ($1.1\%$) of signal from $b$-hadron decay. The cross-section per target nucleon, extrapolated to the full phase-space using \pythia 8 with a specific \lhcb tuning and with the CT09MCS parton distribution functions \cite{CT09MCS}, assuming forward-backward symmetry in the rapidity distribution, thus neglecting a possible small asymmetry induced by nuclear Parton Distribution Functions, is
\begin{equation*}
    \sigma_{\Dz}^{4\pi} =  97.6 \pm 0.7  \stat \pm  9.1  \syst \, \upmu \textrm{b/nucleon}.
\end{equation*}
\begin{figure}[t]
\begin{center}
\includegraphics[width=0.49\textwidth]{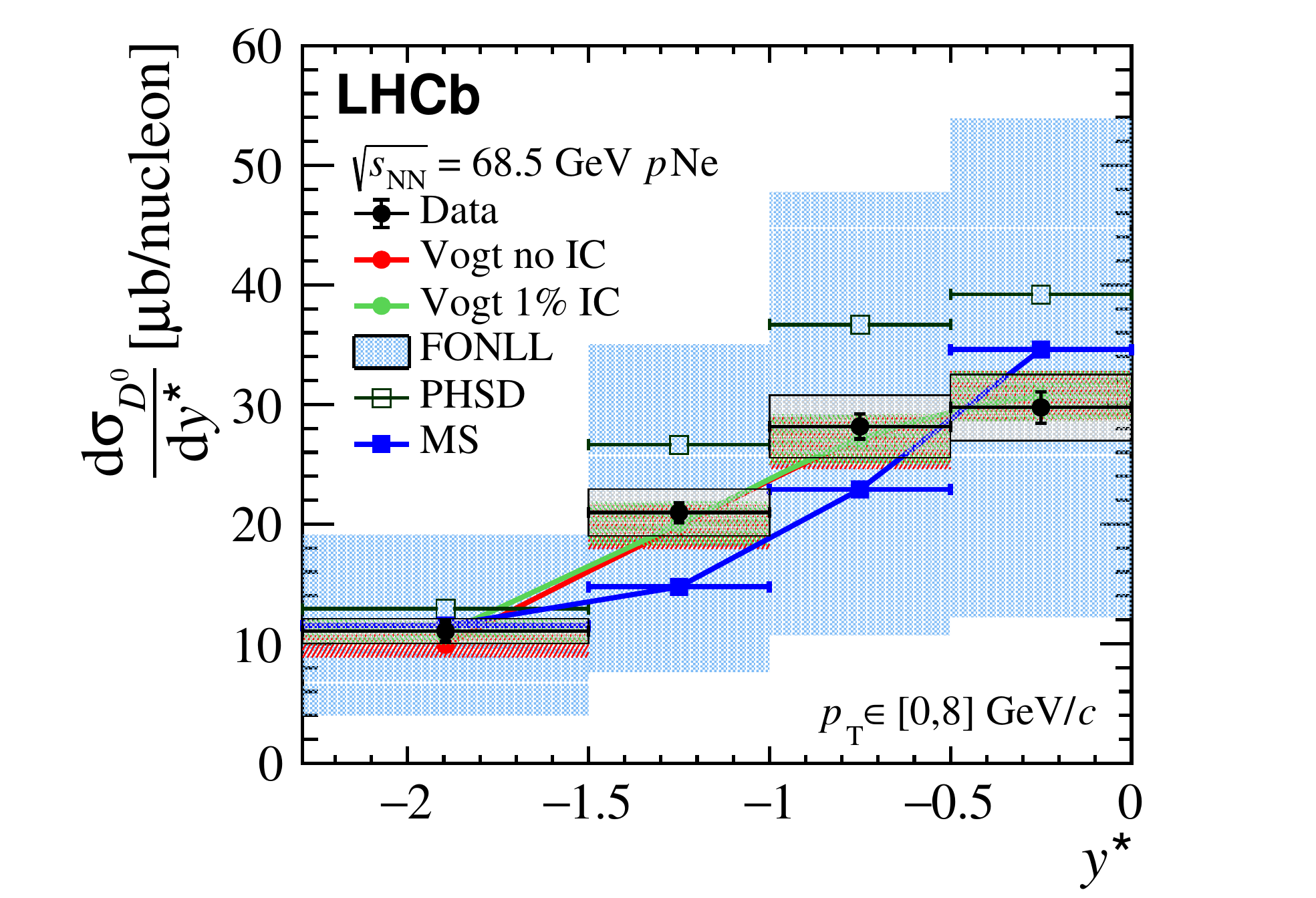}
\includegraphics[width=0.49\textwidth]{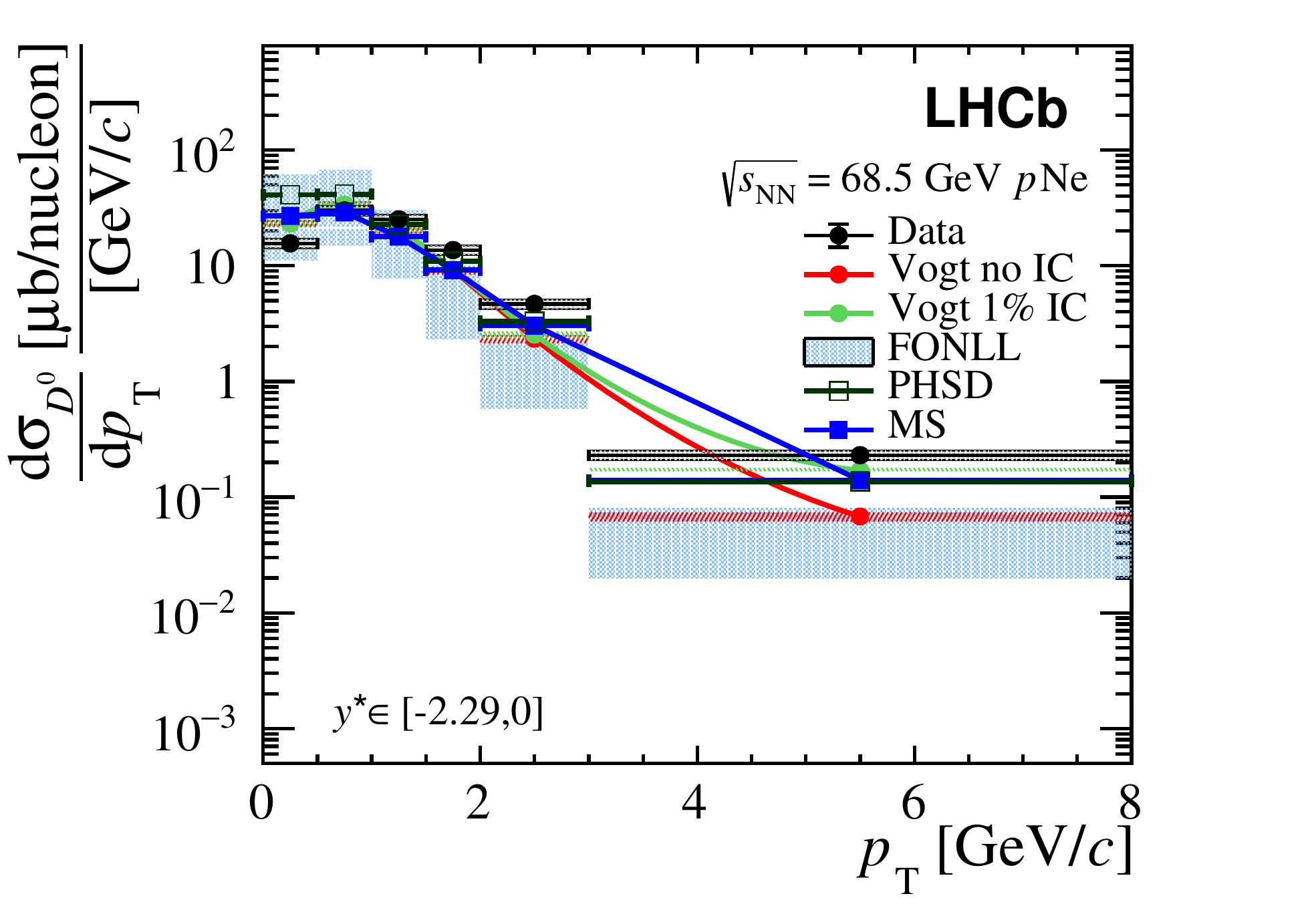}
\caption{Measured \Dz cross-section in $\sqsnn = 68.5\gev$ \pNe collisions as function of (left) $y^\star$ and (right) $\pt$. The quadratic sums of statistical and uncorrelated systematic uncertainties are given by the bars, while the grey boxes represent the correlated systematic uncertainties. The data are overlaid with theoretical predictions as described in the text.}
\label{fig::D0diffcross-section}
\end{center}
\end{figure}
The \Dz  differential cross-sections per target nucleon, as functions of $y^\star$ and $\pt$, are shown in Fig.~\ref{fig::D0diffcross-section}. These results are compared with the fixed-order plus next-to-leading-logarithms resummation model calculations (FONLL) \cite{FONLL1, FONLL2},    and a parton-hadron-string dynamics (PHSD) transport calculation \cite{PHSD}. Both predictions fail to reproduce either the low (PHSD) or high (FONLL) \pt region, while the rapidity shapes are in better agreement with the data. Most features of the LHCb measurements are well described by alternative predictions with (Vogt 1$\%$ IC) or without (Vogt no IC) intrinsic charm contributions, both taking into account the shadowing effect \cite{Vogt}, and by predictions (MS) including 1$\%$ intrinsic charm and 10$\%$ recombination contributions \cite{MS}. Note that only the FONLL predictions include the factorization scale and parton distribution function uncertainties, while the bands in Vogt's predictions represent the variation due to the shadowing, and PHSD or MS models have no uncertainty.
Note also that \Dz and \Dstar measurements have been performed by the STAR experiment in $\sqs=200\gev$ \pp collisions at midrapidity \cite{STAR}, showing a better agreement with FONLL predictions at moderate and large \pt.

\begin{figure}[t]
\begin{center}
\includegraphics[width=0.49\textwidth]{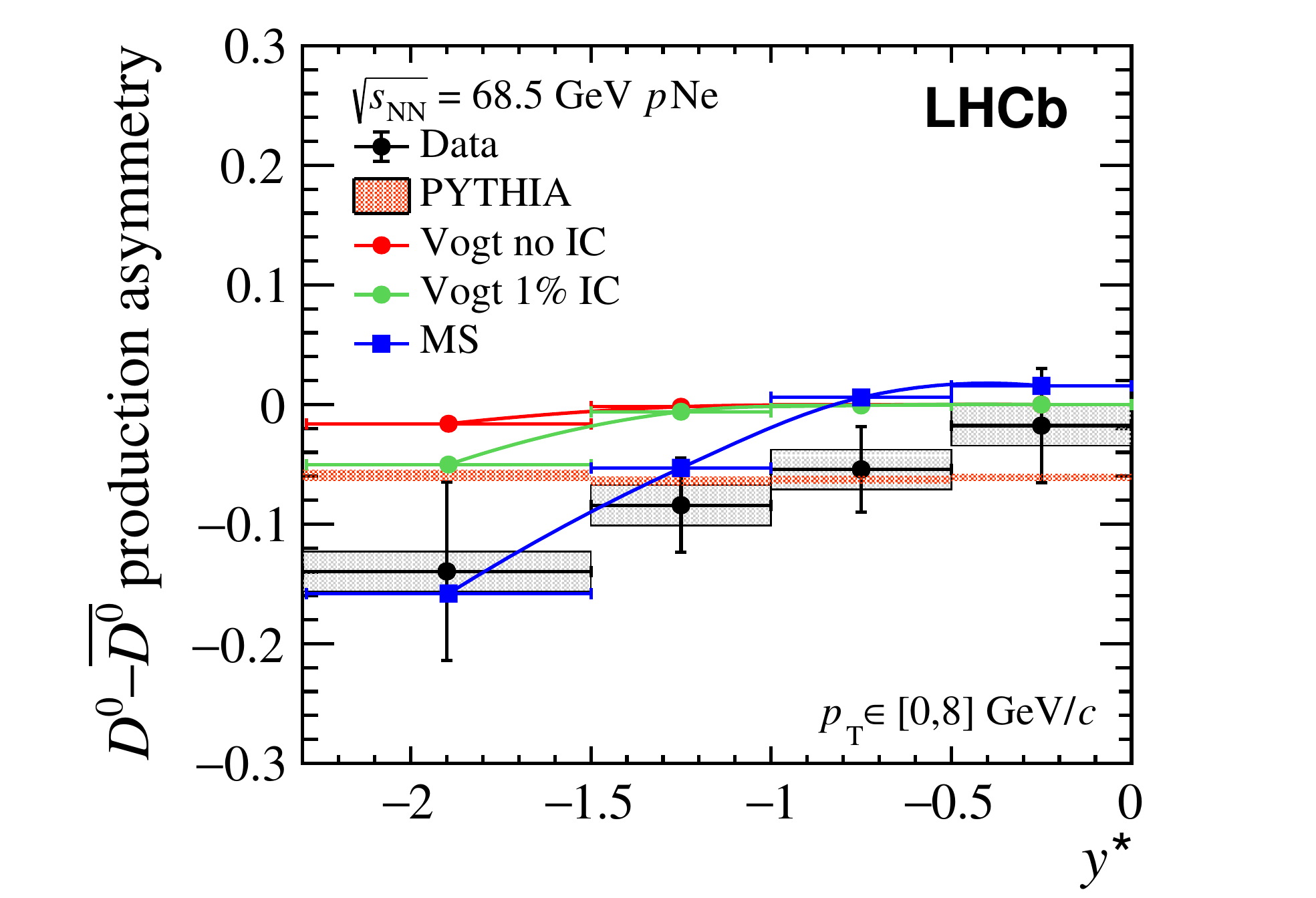}
\includegraphics[width=0.49\textwidth]{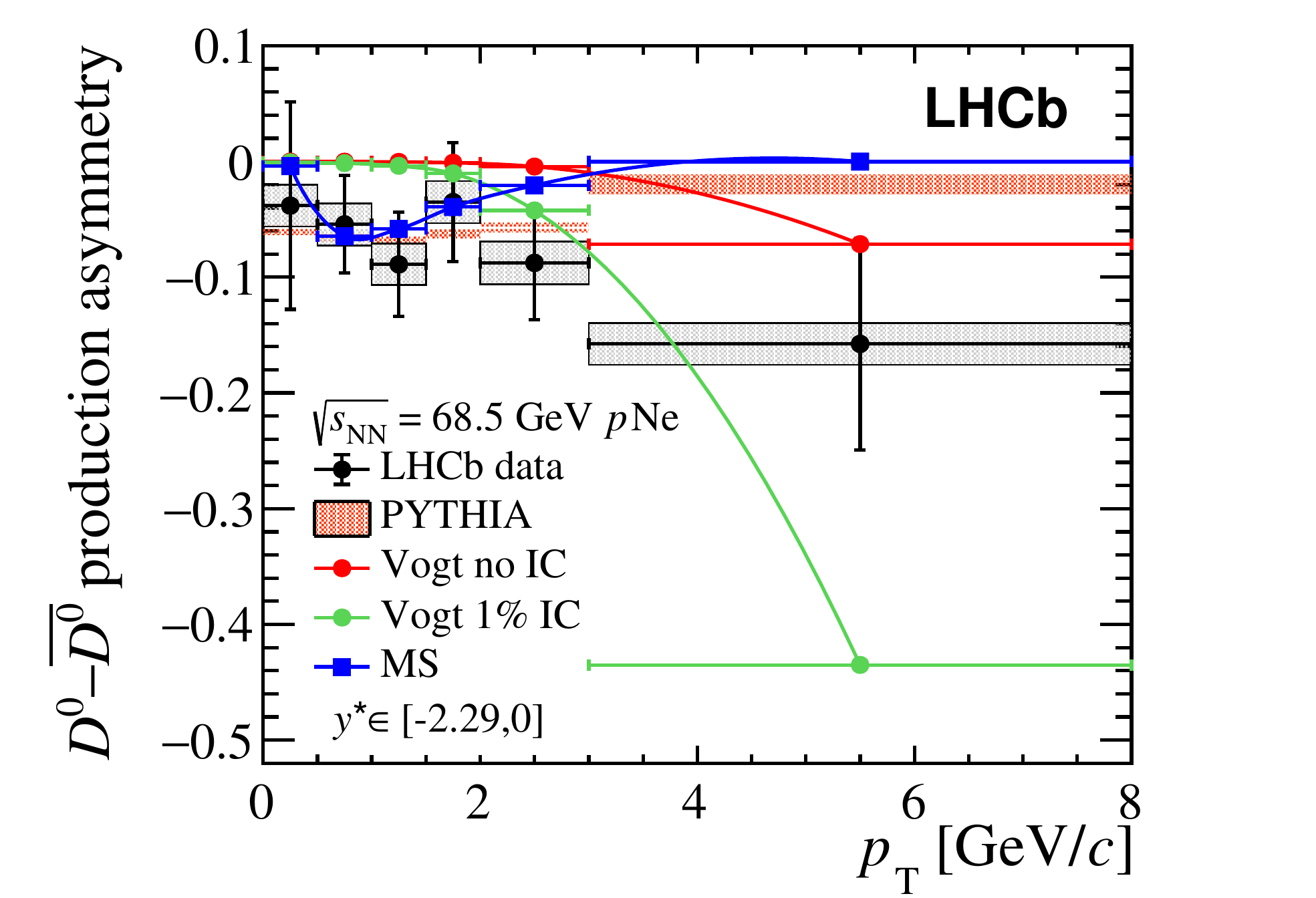}
\caption{The $\Dz-\Dzb$ production asymmetry in $\sqsnn = 68.5\gev$ \pNe collisions as function of (left) centre-of-mass rapidity  $y^\star$, and as function of (right) transverse momentum $\pt$. Quadratic sums of statistical and uncorrelated uncertainties are given by the error bars, while the grey boxes represent the correlated systematic uncertainties. The data are overlaid with theoretical predictions as described in the text.} 
\label{fig::asymmetry}
\end{center}
\end{figure}
Finally, the $\Dz-\Dzb$ production asymmetry is presented. It is quantified as
\begin{equation}
    \mathcal{A}_{\textrm{prod}} = \frac{Y_{\textrm{corr}}(\Dz) - Y_{\textrm{corr}}(\Dzb)}{Y_{\textrm{corr}}(\Dz) + Y_{\textrm{corr}}(\Dzb)},
\end{equation}
where $Y_{\textrm{corr}}(\Dz)$ and $Y_{\textrm{corr}}(\Dzb)$ correspond to the $\Dz\to\Km\pip$ and $\Dzb\to\Kp\pim$ efficiency corrected yields respectively. 
The uncertainties associated with the luminosity determination, neon purity and particle identification efficiency cancel in the asymmetry. Furthermore, uncertainties related to the tracking efficiency and simulation samples partially  cancel.
A conservative systematic uncertainty of $2.8 \%$ is assigned to account for the uncertainty on the simulated material budget of the detector.

The results are presented in Fig. \ref{fig::asymmetry} and indicate a negative asymmetry from $\sim 0$ down to $\sim -15\%$ from $y^\star=0$ to $y^\star=-2.29$. 
The largest asymmetry is obtained at $y^\star=-2.29$, where the valence quark contribution of the neon target is more significant than at $y^\star \sim 0$.
The data are compared with \pythia~8 predictions which show a $\Dz-\Dzb$ asymmetry of about $ -6\%$. This asymmetry, which may be caused by the fragmentation models included in \pythia \cite{Frag98, Frag00}, is compatible with the data, but  mostly independent of $y^\star$ and \pt. While the Vogt predictions from Ref.~\cite{Vogt}, which represent an upper limit,  do not perfectly reproduce the asymmetry scale, they indicate trends compatible with the data for both the rapidity and transverse momentum ratios. 
The MS predictions including 1$\%$ IC and 10$\%$ recombination contributions \cite{MS} are also generally consistent with the data.  

In summary, a study of \Dz meson production in $\sqsnn=68.5\gev$ \pNe collisions with the \lhcb experiment is presented. The sum of the \Dz and \Dzb production cross-section per target nucleon, measured in the centre-of-mass rapidity range $y^\star\in[-2.29,0]$, is found to be $\sigma_{\Dz}^{y^\star\in[-2.29, 0]} =  48.2 \pm 0.3 \stat \pm  4.5  \syst\,\upmu\textrm{b/nucleon}$. The $\Dz-\Dzb$ production asymmetry tends towards a negative value of roughly $-15\%$ in the $y^\star \sim -2$ region, where the valence quark contribution of the neon target is more significant than at $y^\star \sim 0$.
\section*{Acknowledgements}
%
%
\noindent We express our gratitude to our colleagues in the CERN
accelerator departments for the excellent performance of the LHC. We
thank the technical and administrative staff at the LHCb
institutes.
We acknowledge support from CERN and from the national agencies:
CAPES, CNPq, FAPERJ and FINEP (Brazil); 
MOST and NSFC (China); 
CNRS/IN2P3 (France); 
BMBF, DFG and MPG (Germany); 
INFN (Italy); 
NWO (Netherlands); 
MNiSW and NCN (Poland); 
MEN/IFA (Romania); 
MICINN (Spain); 
SNSF and SER (Switzerland); 
NASU (Ukraine); 
STFC (United Kingdom); 
DOE NP and NSF (USA).
We acknowledge the computing resources that are provided by CERN, IN2P3
(France), KIT and DESY (Germany), INFN (Italy), SURF (Netherlands),
PIC (Spain), GridPP (United Kingdom), 
CSCS (Switzerland), IFIN-HH (Romania), CBPF (Brazil),
Polish WLCG  (Poland) and NERSC (USA).
We are indebted to the communities behind the multiple open-source
software packages on which we depend.
Individual groups or members have received support from
ARC and ARDC (Australia);
Minciencias (Colombia);
AvH Foundation (Germany);
EPLANET, Marie Sk\l{}odowska-Curie Actions and ERC (European Union);
A*MIDEX, ANR, IPhU and GLUODYNAMICS/Labex P2IO, and R\'{e}gion Auvergne-Rh\^{o}ne-Alpes (France);
Key Research Program of Frontier Sciences of CAS, CAS PIFI, CAS CCEPP, 
Fundamental Research Funds for the Central Universities, 
and Sci. \& Tech. Program of Guangzhou (China);
GVA, XuntaGal, GENCAT and Prog.~Atracci\'on Talento, CM (Spain);
SRC (Sweden);
the Leverhulme Trust, the Royal Society
 and UKRI (United Kingdom).

\clearpage 

\addcontentsline{toc}{section}{References}
\setboolean{inbibliography}{true}
\ifx\mcitethebibliography\mciteundefinedmacro
\PackageError{LHCb.bst}{mciteplus.sty has not been loaded}
{This bibstyle requires the use of the mciteplus package.}\fi
\providecommand{\href}[2]{#2}

\clearpage
\centerline
{\large\bf LHCb collaboration}
\begin
{flushleft}
\small
R.~Aaij$^{32}$\lhcborcid{0000-0003-0533-1952},
A.S.W.~Abdelmotteleb$^{50}$\lhcborcid{0000-0001-7905-0542},
C.~Abellan~Beteta$^{44}$,
F.~Abudin{\'e}n$^{50}$\lhcborcid{0000-0002-6737-3528},
T.~Ackernley$^{54}$\lhcborcid{0000-0002-5951-3498},
B.~Adeva$^{40}$\lhcborcid{0000-0001-9756-3712},
M.~Adinolfi$^{48}$\lhcborcid{0000-0002-1326-1264},
P.~Adlarson$^{77}$\lhcborcid{0000-0001-6280-3851},
H.~Afsharnia$^{9}$,
C.~Agapopoulou$^{13}$\lhcborcid{0000-0002-2368-0147}
% ...

{\footnotesize \it

% ...
$^{9}$Universit{\'e} Clermont Auvergne, CNRS/IN2P3, LPC, Clermont-Ferrand, France\\
% ...
$^{13}$LPNHE, Sorbonne Universit{\'e}, Paris Diderot Sorbonne Paris Cit{\'e}, CNRS/IN2P3, Paris, France\\
% ...
$^{32}$Nikhef National Institute for Subatomic Physics, Amsterdam, Netherlands\\
% ...
$^{40}$Instituto Galego de F{\'\i}sica de Altas Enerx{\'\i}as (IGFAE), Universidade de Santiago de Compostela, Santiago de Compostela, Spain\\
% ...
$^{44}$Physik-Institut, Universit{\"a}t Z{\"u}rich, Z{\"u}rich, Switzerland\\
% ...
$^{48}$H.H. Wills Physics Laboratory, University of Bristol, Bristol, United Kingdom\\
% ...
$^{50}$Department of Physics, University of Warwick, Coventry, United Kingdom\\
% ...
$^{53}$School of Physics and Astronomy, University of Glasgow, Glasgow, United Kingdom\\
$^{54}$Oliver Lodge Laboratory, University of Liverpool, Liverpool, United Kingdom\\
% ...
$^{77}$Department of Physics and Astronomy, Uppsala University, Uppsala, Sweden, associated to $^{53}$\\
% ...
}
\end{flushleft}


\begin{mcitethebibliography}{10}
\mciteSetBstSublistMode{n}
\mciteSetBstMaxWidthForm{subitem}{\alph{mcitesubitemcount})}
\mciteSetBstSublistLabelBeginEnd{\mcitemaxwidthsubitemform\space}
{\relax}{\relax}

\bibitem{Satz_1994}
H.~Satz and K.~Sridhar, \ifthenelse{\boolean{articletitles}}{\emph{{Charmonium
  production versus open charm in nuclear collisions}},
  }{}\href{http://dx.doi.org/10.1103/PhysRevD.50.3557}{Phys.\ Rev.\
  \textbf{D50} (1994) 3557}\relax
\mciteBstWouldAddEndPuncttrue
\mciteSetBstMidEndSepPunct{\mcitedefaultmidpunct}
{\mcitedefaultendpunct}{\mcitedefaultseppunct}\relax
\EndOfBibitem
\bibitem{maciula2021impact}
R.~Maciula and A.~Szczurek, \ifthenelse{\boolean{articletitles}}{\emph{Impact
  of the \lhcb~{$p\!+\!\!^{4}\!H\!e$} fixed-target ${D^0}/\bar{D^0}$ data on
  the intrinsic \ccbar component in the nucleon},
  }{}\href{http://dx.doi.org/10.1103/PhysRevD.105.014001}{Phys.\ Rev.\
  \textbf{D105} (2022) 014001}\relax
\mciteBstWouldAddEndPuncttrue
\mciteSetBstMidEndSepPunct{\mcitedefaultmidpunct}
{\mcitedefaultendpunct}{\mcitedefaultseppunct}\relax
\EndOfBibitem
\bibitem{D0pHeLHCb}
\lhcb collaboration, R.~Aaij {\em et~al.},
  \ifthenelse{\boolean{articletitles}}{\emph{First measurement of charm
  production in fixed-target configuration at the {LHC}},
  }{}\href{http://dx.doi.org/10.1103/PhysRevLett.122.132002}{Phys.\ Rev.\
  Lett.\  \textbf{102} (2019) 132002}\relax
\mciteBstWouldAddEndPuncttrue
\mciteSetBstMidEndSepPunct{\mcitedefaultmidpunct}
{\mcitedefaultendpunct}{\mcitedefaultseppunct}\relax
\EndOfBibitem
\bibitem{Z0jetLHCb}
\lhcb collaboration, R.~Aaij {\em et~al.},
  \ifthenelse{\boolean{articletitles}}{\emph{Study of {$Z$} bosons produced in
  association with charm in the forward region},
  }{}\href{http://dx.doi.org/10.1103/PhysRevLett.128.082001}{Phys.\ Rev.\
  Lett.\  \textbf{128} (2022) 082001}\relax
\mciteBstWouldAddEndPuncttrue
\mciteSetBstMidEndSepPunct{\mcitedefaultmidpunct}
{\mcitedefaultendpunct}{\mcitedefaultseppunct}\relax
\EndOfBibitem
\bibitem{NNPDF}
The NNPDF Collaboration, R.D.~Ball {\em et~al.},
  \ifthenelse{\boolean{articletitles}}{\emph{Evidence for intrinsic charm quarks in the proton},
  }{}\href{https://doi.org/10.1038/s41586-022-04998-2}{Nature\  \textbf{608} (2022) 483-487}\relax
\mciteBstWouldAddEndPuncttrue
\mciteSetBstMidEndSepPunct{\mcitedefaultmidpunct}
{\mcitedefaultendpunct}{\mcitedefaultseppunct}\relax
\EndOfBibitem
\bibitem{MS}
R.~Maciula and A.~Szczurek,
  \ifthenelse{\boolean{articletitles}}{\emph{Recombination mechanism for
  ${{\ensuremath{\D^0}}\xspace}$-meson production and
  ${{\ensuremath{\D^0}}\xspace}-{{\ensuremath{\Dbar{}^0}}\xspace}$ production
  asymmetry in the {LHCb} {$p\!+\!\!^{20}\!N\!e$} fixed-target experiment},
  }{}\href{http://dx.doi.org/10.1016/j.physletb.2022.137530}{Phys.\ Lett.\
  \textbf{B835} (2022) 137530},
  \href{http://arxiv.org/abs/2206.02750}{{\normalfont\ttfamily
  arXiv:2206.02750}}\relax
\mciteBstWouldAddEndPuncttrue
\mciteSetBstMidEndSepPunct{\mcitedefaultmidpunct}
{\mcitedefaultendpunct}{\mcitedefaultseppunct}\relax
\EndOfBibitem
\bibitem{E791}
Fermilab E791 Collaboration, E.~M.~Aitala {\em et~al.}, \ifthenelse{\boolean{articletitles}}{\emph{Asymetries between the production of $D^+$ and $D^-$ mesons from 500 GeV/$c$ $\pi^-$-nucleus interactions as a function of $x_F$ and $p_t^2$},
  }{}\href{https://doi.org/10.1016/0370-2693(96)00093-7}{Phys.\ Lett.\
  \textbf{B371} (1996) 157-162}\relax
\mciteBstWouldAddEndPuncttrue
\mciteSetBstMidEndSepPunct{\mcitedefaultmidpunct}
{\mcitedefaultendpunct}{\mcitedefaultseppunct}\relax
\EndOfBibitem
\bibitem{Braaten}
E. Braaten, Y. Jia, and T. Mehen, \ifthenelse{\boolean{articletitles}}{\emph{Charm-anticharm asymetries in photoproduction from heavy-quark recombination},
  }{}\href{https://doi-org.ezproxy.cern.ch/10.1103/PhysRevD.66.014003}{Phys.\ Rev.\ D
  \textbf{66} (2002) 014003}\relax
\mciteBstWouldAddEndPuncttrue
\mciteSetBstMidEndSepPunct{\mcitedefaultmidpunct}
{\mcitedefaultendpunct}{\mcitedefaultseppunct}\relax
\EndOfBibitem
\bibitem{Alves:2008zz}
LHCb collaboration, A.~A. Alves~Jr.\ {\em et~al.},
  \ifthenelse{\boolean{articletitles}}{\emph{{The \lhcb detector at the LHC}},
  }{}\href{http://dx.doi.org/10.1088/1748-0221/3/08/S08005}{JINST \textbf{3}
  (2008) S08005}\relax
\mciteBstWouldAddEndPuncttrue
\mciteSetBstMidEndSepPunct{\mcitedefaultmidpunct}
{\mcitedefaultendpunct}{\mcitedefaultseppunct}\relax
\EndOfBibitem
\bibitem{LHCb-DP-2014-002}
LHCb collaboration, R.~Aaij {\em et~al.},
  \ifthenelse{\boolean{articletitles}}{\emph{{LHCb detector performance}},
  }{}\href{http://dx.doi.org/10.1142/S0217751X15300227}{Int.\ J.\ Mod.\ Phys.\
  \textbf{A30} (2015) 1530022},
  \href{http://arxiv.org/abs/1412.6352}{{\normalfont\ttfamily
  arXiv:1412.6352}}\relax
\mciteBstWouldAddEndPuncttrue
\mciteSetBstMidEndSepPunct{\mcitedefaultmidpunct}
{\mcitedefaultendpunct}{\mcitedefaultseppunct}\relax
\EndOfBibitem
\bibitem{LHCb-PAPER-2014-047}
LHCb collaboration, R.~Aaij {\em et~al.},
  \ifthenelse{\boolean{articletitles}}{\emph{{Precision luminosity measurements
  at LHCb}}, }{}\href{http://dx.doi.org/10.1088/1748-0221/9/12/P12005}{JINST
  \textbf{9} (2014) P12005},
  \href{http://arxiv.org/abs/1410.0149}{{\normalfont\ttfamily
  arXiv:1410.0149}}\relax
\mciteBstWouldAddEndPuncttrue
\mciteSetBstMidEndSepPunct{\mcitedefaultmidpunct}
{\mcitedefaultendpunct}{\mcitedefaultseppunct}\relax
\EndOfBibitem
\bibitem{trigger}
LHCb collaboration, R.~Aaij {\em et~al.},
  \ifthenelse{\boolean{articletitles}}{\emph{{The LHCb trigger and its
  performance in 2011}},
  }{}\href{http://dx.doi.org/10.1088/1748-0221/8/04/P04022}{JINST \textbf{8}
  (2013) P04022}, \href{http://arxiv.org/abs/1211.3055}{{\normalfont\ttfamily
  arXiv:1211.3055}}\relax
\mciteBstWouldAddEndPuncttrue
\mciteSetBstMidEndSepPunct{\mcitedefaultmidpunct}
{\mcitedefaultendpunct}{\mcitedefaultseppunct}\relax
\EndOfBibitem
\bibitem{LHCb-PAPER-2018-023}
LHCb collaboration, R.~Aaij {\em et~al.},
  \ifthenelse{\boolean{articletitles}}{\emph{{First measurement of charm
  production in fixed-target configuration at the LHC}},
  }{}\href{http://dx.doi.org/10.1103/PhysRevLett.122.132002}{Phys.\ Rev.\
  Lett.\  \textbf{122} (2019) 132002},
  \href{http://arxiv.org/abs/1810.07907}{{\normalfont\ttfamily
  arXiv:1810.07907}}\relax
\mciteBstWouldAddEndPuncttrue
\mciteSetBstMidEndSepPunct{\mcitedefaultmidpunct}
{\mcitedefaultendpunct}{\mcitedefaultseppunct}\relax
\EndOfBibitem
\bibitem{Pythia8}
T.~Sj{\"o}strand, S.~Mrenna, and P.~Skands,
  \ifthenelse{\boolean{articletitles}}{\emph{{A brief introduction to \pythia
  8.1}}, }{}\href{http://dx.doi.org/10.1016/j.cpc.2008.01.036}{Comput.\ Phys.\
  Commun.\  \textbf{178} (2008) 852},
  \href{http://arxiv.org/abs/0710.3820}{{\normalfont\ttfamily
  arXiv:0710.3820}}\relax
\mciteBstWouldAddEndPuncttrue
\mciteSetBstMidEndSepPunct{\mcitedefaultmidpunct}
{\mcitedefaultendpunct}{\mcitedefaultseppunct}\relax
\EndOfBibitem
\bibitem{Lund}
B. Andersson, G. Gutafson, G. Ingelman and T.~Sj{\"o}strand, \ifthenelse{\boolean{articletitles}}{\emph{{Parton fragmentation and string dynamics}},
  }{}\href{https://doi.org/10.1016/0370-1573(83)90080-7}{Physics Report
  \textbf{97} (1983) 31-45}\relax
\mciteBstWouldAddEndPuncttrue
\mciteSetBstMidEndSepPunct{\mcitedefaultmidpunct}
{\mcitedefaultendpunct}{\mcitedefaultseppunct}\relax
\EndOfBibitem
\bibitem{1742-6596-331-3-032047}
LHCb collaboration, I.~Belyaev {\em et~al.},
  \ifthenelse{\boolean{articletitles}}{\emph{{Handling of the generation of
  primary events in Gauss, the LHCb simulation framework}},
  }{}\href{http://dx.doi.org/10.1088/1742-6596/331/3/032047}{J.\ Phys.\ Conf.\
  Ser.\  \textbf{331} (2011) 032047}\relax
\mciteBstWouldAddEndPuncttrue
\mciteSetBstMidEndSepPunct{\mcitedefaultmidpunct}
{\mcitedefaultendpunct}{\mcitedefaultseppunct}\relax
\EndOfBibitem
\bibitem{EvtGen}
D.~J. Lange, \ifthenelse{\boolean{articletitles}}{\emph{{The EvtGen particle
  decay simulation package}},
  }{}\href{http://dx.doi.org/10.1016/S0168-9002(01)00089-4}{Nucl.\ Instrum.\
  Meth.\  \textbf{A462} (2001) 152}\relax
\mciteBstWouldAddEndPuncttrue
\mciteSetBstMidEndSepPunct{\mcitedefaultmidpunct}
{\mcitedefaultendpunct}{\mcitedefaultseppunct}\relax
\EndOfBibitem
\bibitem{PHOTOS}
P.~Golonka and Z.~Was, \ifthenelse{\boolean{articletitles}}{\emph{{PHOTOS Monte
  Carlo: a precision tool for QED corrections in $Z$ and $W$ decays}},
  }{}\href{http://dx.doi.org/10.1140/epjc/s2005-02396-4}{Eur.\ Phys.\ J.\
  \textbf{C45} (2006) 97}\relax
\mciteBstWouldAddEndPuncttrue
\mciteSetBstMidEndSepPunct{\mcitedefaultmidpunct}
{\mcitedefaultendpunct}{\mcitedefaultseppunct}\relax
\EndOfBibitem
\bibitem{Pierog:2013ria}
T.~Pierog {\em et~al.}, \ifthenelse{\boolean{articletitles}}{\emph{{EPOS LHC:
  Test of collective hadronization with data measured at the CERN Large Hadron
  Collider}}, }{}\href{http://dx.doi.org/10.1103/PhysRevC.92.034906}{Phys.\
  Rev.\  \textbf{C92} (2015) 034906},
  \href{http://arxiv.org/abs/1306.0121}{{\normalfont\ttfamily
  arXiv:1306.0121}}\relax
\mciteBstWouldAddEndPuncttrue
\mciteSetBstMidEndSepPunct{\mcitedefaultmidpunct}
{\mcitedefaultendpunct}{\mcitedefaultseppunct}\relax
\EndOfBibitem
\bibitem{geant2}
\geant collaboration, J.~Allison {\em et~al.},
  \ifthenelse{\boolean{articletitles}}{\emph{\geant developments and
  applications}, }{}\href{http://dx.doi.org/10.1109/TNS.2006.869826}{IEEE
  Trans.\ Nucl.\ Sci.\  \textbf{53} (2006) 270}\relax
\mciteBstWouldAddEndPuncttrue
\mciteSetBstMidEndSepPunct{\mcitedefaultmidpunct}
{\mcitedefaultendpunct}{\mcitedefaultseppunct}\relax
\EndOfBibitem
\bibitem{geant4}
\geant collaboration, S.~Agostinelli {\em et~al.},
  \ifthenelse{\boolean{articletitles}}{\emph{\geant: a simulation toolkit},
  }{}\href{http://dx.doi.org/10.1016/S0168-9002(03)01368-8}{Nucl.\ Instr.\
  Meth.\  \textbf{506} (2003) 250}\relax
\mciteBstWouldAddEndPuncttrue
\mciteSetBstMidEndSepPunct{\mcitedefaultmidpunct}
{\mcitedefaultendpunct}{\mcitedefaultseppunct}\relax
\EndOfBibitem
\bibitem{gauss}
M.~Clemencic {\em et~al.}, \ifthenelse{\boolean{articletitles}}{\emph{{The LHCb
  simulation application, Gauss: Design, evolution and experience}},
  }{}\href{http://dx.doi.org/10.1088/1742-6596/331/3/032023}{J.\ Phys.\ Conf.\
  Ser.\  \textbf{331} (2011) 032023}\relax
\mciteBstWouldAddEndPuncttrue
\mciteSetBstMidEndSepPunct{\mcitedefaultmidpunct}
{\mcitedefaultendpunct}{\mcitedefaultseppunct}\relax
\EndOfBibitem
\bibitem{LHCb-PUB-2016-021}
L.~Anderlini {\em et~al.}, \ifthenelse{\boolean{articletitles}}{\emph{{The
  PIDCalib package}}, }{}
  \href{http://cdsweb.cern.ch/search?p=LHCb-PUB-2016-021&f=reportnumber&action_search=Search&c=LHCb+Notes}
  {LHCb-PUB-2016-021}\relax
\mciteBstWouldAddEndPuncttrue
\mciteSetBstMidEndSepPunct{\mcitedefaultmidpunct}
{\mcitedefaultendpunct}{\mcitedefaultseppunct}\relax
\EndOfBibitem
\bibitem{tracking}
LHCb collaboration, R.~Aaij {\em et~al.},
  \ifthenelse{\boolean{articletitles}}{\emph{{Measurement of the track reconstruction efficiency at LHCb}},
  }{}\href{https://dx.doi.org/10.1088/1748-0221/10/02/P02007}{JINST  \textbf{10} (2015) P02007}\relax
\mciteBstWouldAddEndPuncttrue
\mciteSetBstMidEndSepPunct{\mcitedefaultmidpunct}
{\mcitedefaultendpunct}{\mcitedefaultseppunct}\relax
\EndOfBibitem
\bibitem{antiproton}
LHCb collaboration, R.~Aaij {\em et~al.},
  \ifthenelse{\boolean{articletitles}}{\emph{{Measurement of antiproton
  production in {$p$He} collisions at $\sqsnn=110$ GeV}},
  }{}\href{http://dx.doi.org/10.1103/PhysRevLett.121.222001}{Phys.\ Rev.\
  Lett.\  \textbf{121} (2018) 222001}\relax
\mciteBstWouldAddEndPuncttrue
\mciteSetBstMidEndSepPunct{\mcitedefaultmidpunct}
{\mcitedefaultendpunct}{\mcitedefaultseppunct}\relax
\EndOfBibitem
\bibitem{pdg}
Particle Data Group, R.~L. Workman {\em et~al.},
  \ifthenelse{\boolean{articletitles}}{\emph{{\href{http://pdg.lbl.gov/}{Review
  of particle physics}}},
  }{}\href{http://dx.doi.org/10.1093/ptep/ptac097}{Prog.\ Theor.\ Exp.\ Phys.\
  \textbf{2022} (2022), no.~8 083C01}\relax
\mciteBstWouldAddEndPuncttrue
\mciteSetBstMidEndSepPunct{\mcitedefaultmidpunct}
{\mcitedefaultendpunct}{\mcitedefaultseppunct}\relax
\EndOfBibitem
\bibitem{CT09MCS}
H.-L. Lai {\em et~al.}, \ifthenelse{\boolean{articletitles}}{\emph{{Parton
  distributions for event generators}},
  }{}\href{http://dx.doi.org/10.1007/JHEP04(2010)035}{JHEP \textbf{04} (2010)
  035}\relax
\mciteBstWouldAddEndPuncttrue
\mciteSetBstMidEndSepPunct{\mcitedefaultmidpunct}
{\mcitedefaultendpunct}{\mcitedefaultseppunct}\relax
\EndOfBibitem
\bibitem{FONLL1}
M.~Cacciari, P.~Nason, and R.~Vogt,
  \ifthenelse{\boolean{articletitles}}{\emph{{QCD} predictions for charm and
  bottom quark production at {RHIC}},
  }{}\href{http://dx.doi.org/10.1103/PhysRevLett.95.122001}{Phys.\ Rev.\ Lett.\
   \textbf{95} (2005) 122001}\relax
\mciteBstWouldAddEndPuncttrue
\mciteSetBstMidEndSepPunct{\mcitedefaultmidpunct}
{\mcitedefaultendpunct}{\mcitedefaultseppunct}\relax
\EndOfBibitem
\bibitem{FONLL2}
M.~Cacciari, M.~Greco, and P.~Nason,
  \ifthenelse{\boolean{articletitles}}{\emph{The \pt spectrum in heavy-flavour
  hadroproduction},
  }{}\href{http://dx.doi.org/10.1088/1126-6708/1998/05/007}{JHEP \textbf{05}
  (1998) 007}\relax
\mciteBstWouldAddEndPuncttrue
\mciteSetBstMidEndSepPunct{\mcitedefaultmidpunct}
{\mcitedefaultendpunct}{\mcitedefaultseppunct}\relax
\EndOfBibitem
\bibitem{PHSD}
T.~Song {\em et~al.}, \ifthenelse{\boolean{articletitles}}{\emph{Single
  electrons from heavy-flavor mesons in relativistic heavy-ion collisions},
  }{}\href{http://dx.doi.org/10.1103/PhysRevC.96.014905}{Phys.\ Rev.\
  \textbf{C96} (2017) 014905}\relax
\mciteBstWouldAddEndPuncttrue
\mciteSetBstMidEndSepPunct{\mcitedefaultmidpunct}
{\mcitedefaultendpunct}{\mcitedefaultseppunct}\relax
\EndOfBibitem
\bibitem{Vogt}
R.~Vogt, \ifthenelse{\boolean{articletitles}}{\emph{Limits on intrinsic charm
  production from the {S}ea{Q}uest experiment},
  }{}\href{http://dx.doi.org/10.1103/PhysRevC.103.035204}{Phys.\ Rev.\
  \textbf{C103} (2021) 035204}\relax
\mciteBstWouldAddEndPuncttrue
\mciteSetBstMidEndSepPunct{\mcitedefaultmidpunct}
{\mcitedefaultendpunct}{\mcitedefaultseppunct}\relax
\EndOfBibitem
\bibitem{STAR}
STAR Collaboration, L.~Adamczyk {\em et~al.}, \ifthenelse{\boolean{articletitles}}{\emph{Measurements of \Dz and \Dstar production in $p+p$} collisions at $\sqs=200\gev$,
  }{}\href{https://doi-org.ezproxy.cern.ch/10.1103/PhysRevD.86.072013}{Phys.\ Rev.\ \textbf{D86} (2012) 072013}\relax
\mciteBstWouldAddEndPuncttrue
\mciteSetBstMidEndSepPunct{\mcitedefaultmidpunct}
{\mcitedefaultendpunct}{\mcitedefaultseppunct}\relax
\EndOfBibitem
\bibitem{Frag98}
E. Norrbin and T.~Sj{\"o}strand, \ifthenelse{\boolean{articletitles}}{\emph{Production mechanisms of charm hadrons in the string model},
  }{}\href{https://doi.org/10.1016/S0370-2693(98)01244-1}{Phys.\ Lett.\
  \textbf{B442} (1998) 407-416}\relax
\mciteBstWouldAddEndPuncttrue
\mciteSetBstMidEndSepPunct{\mcitedefaultmidpunct}
{\mcitedefaultendpunct}{\mcitedefaultseppunct}\relax
\EndOfBibitem
\bibitem{Frag00}
E. Norrbin and T.~Sj{\"o}strand, \ifthenelse{\boolean{articletitles}}{\emph{Production and hadronization of heavy quarks},
  }{}\href{https://doi-org.ezproxy.cern.ch/10.1007/s100520000460}{Eur.\ Phys.\ J
  \textbf{C17} (2000) 137-161}\relax
\mciteBstWouldAddEndPuncttrue
\mciteSetBstMidEndSepPunct{\mcitedefaultmidpunct}
{\mcitedefaultendpunct}{\mcitedefaultseppunct}\relax
\EndOfBibitem
\end{mcitethebibliography}
\end{document}